\documentclass[showpacs,prb,twocolumn]{revtex4}
\usepackage{amsfonts}
\usepackage{amsmath}
\usepackage{amssymb}
\usepackage{graphicx}

\begin{document}

\title{Spin and orbital valence bond solids in a one-dimensional
spin-orbital system: Schwinger boson mean field theory}
\author{Peng Li and Shun-Qing Shen}
\affiliation{Department of Physics, and Center for Theoretical and Computational Physics,
The University of Hong Kong, Pokfulam Road, Hong Kong, China}

\begin{abstract}
A generalized one-dimensional $SU(2)\times SU(2)$ spin-orbital model is
studied by Schwinger boson mean-field theory (SBMFT). We explore mainly the
dimer phases and clarify how to capture properly the low temperature
properties of such a system by SBMFT. The phase diagrams are exemplified.
The three dimer phases, orbital valence bond solid (OVB) state, spin valence
bond solid (SVB) state and spin-orbital valence bond solid (SOVB) state, are
found to be favored in respectively proper parameter regions, and they can
be characterized by the static spin and pseudospin susceptibilities
calculated in SBMFT scheme. The result reveals that the spin-orbit coupling
of $SU(2)\times SU(2)$\ type serves as both the spin-Peierls and
orbital-Peierles mechanisms that responsible for the spin-singlet and
orbital-singlet formations respectively.
\end{abstract}

\date{\today}
\maketitle

\section{Introduction}

Looking for exotic states in transition-metal oxides is a fascinating
problem. Spin-orbital models arose from the consideration of orbital degrees
of freedom of $d$- or $f$-electron in transition metals.\cite{TN,KK,R} When
Hund's coupling, orbital anisotropy, Jahn-Teller effect, etc. are ignored, a
$SU(2)\times SU(2)$ spin-orbital model is recurrently proposed and studied
in connection with real materials, such as C$_{60}$ material,\cite{AA}
spin-gap materials Na$_{2}$Ti$_{2}$Sb$_{2}$O and NaV$_{2}$O$_{5}$,\cite%
{PSK,Fujii} and cubic vanadates LaVO$_{3}$ and YVO$_{3}$.\cite{KHO} In this
paper, we employ the Schwinger boson mean-field theory (SBMFT) to study the
low temperature properties of the generalized one-dimensional ($1D$) $%
SU(2)\times SU(2)$ spin-orbital model and reveal that the dimer phases in
fact consist of three kinds of valence bond phases: orbital valence bond
solid (OVB) state, spin valence bond solid (SVB) state and spin-orbital
valence bond solid (SOVB) state. All of the three phases are gapped, but
they can be well characterized by the spin and pseudospin susceptibilities,
and thus are distinguishable experimentally. As we will reveal, the $%
SU(2)\times SU(2)$ type of spin-orbit coupling is responsible for both the
spin-Peierls and orbital-Peierls phenomena occurring in this $1D$ model.

This paper is organized as follows. In Section II, we give a detailed
Schwinger boson mean-field scheme on the spin-orbital model. Then in Section
III, we study the phase diagrams for two cases: (i) $S=1$ and $T=1/2$; (ii) $%
S=1/2$ and $T=1/2$. The phase diagram for the former is novel while the one
for the latter is a supplementary for previous works.\cite{PSK,I,ZO} In
Section IV, we deduce the spin and pseudospin susceptibilities. At last, a
brief conclusion is given in Section V.

\section{Model Hamiltonian and Schwinger boson mean-field theory}

The one-dimensional spin-orbital Hamiltonian with the $SU(2)\times SU(2)$
symmetry reads%
\begin{equation}
H=\sum\limits_{m}\left( \mathbf{S}_{m}\cdot\mathbf{S}_{m+1}+x\right) \left(
\mathbf{T}_{m}\cdot\mathbf{T}_{m+1}+y\right) ,  \label{h}
\end{equation}
where $\mathbf{S}$ and $\mathbf{T}$ are two sets of spin operators which
satisfy the $SU(2)$ algebra with eigenvalues $S$ and $T$, respectively, and $%
x$ and $y$ are two constant parameters. It can also be viewed as a
generalized spin ladder system with four-operator interactions. There are
two special points in this model. The first is the \textit{dimer point} $D$:
$\left( x,y\right) =\left( S(S+1),T(T+1)\right) $, whose ground state is
perfectly dimerized and can be recognized as a two-fold degenerate dimerized
valence bond solid.\cite{K} Another point is the \textit{FM point} $F$:\ $%
\left( x,y\right) =\left( -S^{2},-T^{2}\right) $, whose ground state
possesses the maximal values for both $\mathbf{S}_{total}=\sum \limits_{m}%
\mathbf{S}_{m}$ and $\mathbf{T}_{total}=\sum\limits_{m}\mathbf{T}_{m}$, and
is, we will see in the phase diagrams later, the critical point of three
uniform phases. The model for $S=1/2$ and $T=1/2$ was studied both
analytically and numerically, and the phase diagrams were proposed by
several authors.\cite{PSK,I,ZO} Except for the phases with conventional spin
or orbital ferromagnetic (FM)\textit{\ }or antiferromagnetic (AFM) orders,
the mixed spin-orbital valence bond state, which exhibits no long-range
correlation and no gap near the $SU(4)$ symmetric point $S$: $\left(
x,y\right) =\left( 1/4,1/4\right) $,\cite{L} appears in a large regime in
the phase diagram. The model for $S=1$ and $T=1/2$ around the point $O$: $%
\left( x,y\right) =\left( 1,1/4\right) $ was revealed to exhibit the OVB
state.\cite{SXZ,SK,U}

In this paper we will employ SBMFT to study the phase diagram, and focus on
exploring the dimer phases. SBMFT had been introduced to the field of
strongly correlated systems for a long time, and shows its merit in
describing the spin systems at low temperatures. It can produce spin liquid
phase and spin gap close to the ground state as well as the phases with
long-range correlation.\cite{A,AA1,S,Zhang01prl} In Schwinger boson
representation, one can introduce two boson operators $a$ and $b$ to
represent a spin operator $\mathbf{S}_{m}$: $\ $%
\begin{equation}
S_{m}^{+}=a_{m}^{\dagger}b_{m},S_{m}^{-}=b_{m}^{\dagger}a_{m},S_{m}^{z}=%
\frac{1}{2}\left( a_{m}^{\dagger}a_{m}-b_{m}^{\dagger}b_{m}\right)
\end{equation}
with the local constraint
\begin{equation}
a_{m}^{\dagger}a_{m}+b_{m}^{\dagger}b_{m}=2S.  \label{constraint}
\end{equation}
In this way one can introduce two types of bond operators,

\begin{subequations}
\begin{align}
F_{mn} & =\frac{1}{2}\left( a_{m}^{\dagger}a_{n}+b_{m}^{\dagger}b_{n}\right)
, \\
A_{mn} & =\frac{1}{2}\left( a_{m}b_{n}-b_{m}a_{n}\right) ,
\end{align}
to mimic the ferromagnetic (FM) and antiferromagnetic (AFM) channels of the
spin $S$ Heisenberg interactions. The low energy physics can be well
captured by the mean fields, $\Delta_{F}=\left\langle F_{mn}\right\rangle $
or $\Delta_{A}=\left\langle A_{mn}\right\rangle $, where $\left\langle
...\right\rangle $ stands for the thermodynamic average. Likewise, the mean
fields for the pseudospin $T$\ are denoted by $\Theta_{F}$ and $\Theta_{A}$
for the FM and AFM channels, respectively.

As revealed by Ceccatto \textit{et al}.,\cite{C} retaining both the FM and
the AFM channels will lead to

\end{subequations}
\begin{equation}
\mathbf{S}_{m}\cdot\mathbf{S}_{n}=\text{:}F_{mn}^{\dagger}F_{mn}\text{:}%
-A_{mn}^{\dagger}A_{mn}  \label{cd2}
\end{equation}
where :$\ $: denotes normal order. In the mean field scheme, the bond operators are decomposed as%
\begin{equation}
\{%
\begin{array}{c}
\text{:}F_{mn}^{\dagger}F_{mn}\text{:}\rightarrow\Delta_{F}(F_{mn}+F_{mn}^{%
\dagger})-\Delta_{F}^{2}, \\
A_{mn}^{\dagger}A_{mn}\rightarrow\Delta_{A}(A_{mn}+A_{mn}^{\dagger})-%
\Delta_{A}^{2},%
\end{array}%
\end{equation}
where the higher-order terms, $\left(
F_{mn}^{\dagger}-\Delta_{F}\right) \left(
F_{mn}-\Delta_{F}\right) $ and $\left( A_{mn}^{\dagger}-\Delta
_{A}\right) \left( A_{mn}-\Delta_{A}\right) $, are ignored. We shall avoid applying the identity,

\begin{equation}
\text{:}F_{mn}^{\dagger}F_{mn}\text{:}+A_{mn}^{\dagger}A_{mn}=S^{2},
\label{Identity}
\end{equation}
which holds due to the constraint Eq. (\ref{constraint}) and implies

\begin{equation}
\mathbf{S}_{m}\cdot\mathbf{S}_{n}=\{%
\begin{array}{c}
2\text{:}F_{mn}^{\dagger}F_{mn}\text{:}-S^{2}\text{;\ for FM,} \\
-2A_{mn}^{\dagger}A_{mn}+S^{2}\text{; for AFM.}%
\end{array}
\label{wd}
\end{equation}
The identity, Eq. (\ref{Identity}), is largely violated when the
constraint is imposed only on average. And this is also true even
when the Gaussian-fluctuation corrections are considered.\cite{T}

To show how Eq. (\ref{cd2}) works, we present a two-site example, $%
H=J\mathbf{S}_{1}\cdot\mathbf{S}_{2}$, which can be solved analytically.
After solving it by SBMFT, one can find, in the FM ($J<0$) case, Eq. (\ref%
{cd2}) gives ground energy $f_{0}=JS^{2}$ with $\Delta_{F}=S$ and $%
\Delta_{A}=0$, and Eq. (\ref{wd}) gives the same ground energy with $%
\Delta_{F}=S$. They both are in agreement with the exact
solution. In the AFM ($J>0$) case, Eq. (\ref{cd2}) gives
$f_{0}=-JS(S+1)$ with $\Delta_{A}=\sqrt{S(S+1)}$ and
$\Delta_{F}=0$, while Eq. (\ref{wd}) gives $f_{0}=-JS(S+2)$ with
$\Delta_{A}=\sqrt{S(S+1)}$. The former is just the exact energy;
the latter approaches the exact value only in the limit $%
S\rightarrow\infty$. Thus Eq. (\ref{wd}) overestimates the ground
energy in AFM channel. We have also found that Eq. (\ref{cd2}),
instead of Eq. (\ref{wd}), can produce the exact energy of the
dimer point in the spin-orbital model of Eq. (\ref{h}). Thus for
a small $S$, the mean field scheme in Eq. (\ref{cd2}) is better
than that in Eq. (\ref{wd}). In practice, when solving the mean
field equations, we found that only one of the FM and AFM channels
can survive on a bond meanwhile for $1D$ and $2D$ unfrustrated
systems.

To perform the mean field calculation, we decompose Eq. (\ref{h}) into the
spin $S$ and the pseudospin $T$ chains,

\begin{align}
H & \approx\sum\limits_{m}[J_{S}(m)\mathbf{S}_{m}\cdot\mathbf{S}%
_{m+1}+J_{T}(m)\mathbf{T}_{m}\cdot\mathbf{T}_{m+1}]  \notag \\
& +\sum\limits_{m}\left[ yJ_{T}(m)+xJ_{S}(m)+J_{S}(m)J_{T}(m)\right]
\label{hm}
\end{align}
where the effective couplings

\begin{subequations}
\label{IJ}
\begin{align}
J_{T}(m)& =\left\langle \mathbf{S}_{m}\cdot \mathbf{S}_{m+1}+x\right\rangle ,
\label{IJa} \\
J_{S}(m)& =\left\langle \mathbf{T}_{m}\cdot \mathbf{T}_{m+1}+y\right\rangle .
\label{IJb}
\end{align}%
The two chains are coupled by the effective couplings, which will
be determined self-consistently. And Eq. (\ref{cd2}) ensures a
reasonable estimation of the strength of the effective couplings,
Eq. (\ref{IJ}), for both of the FM and AFM channels. One should
notice, for $S=T=1/2$, the mixed channel of $S$ and $T$ must be
considered around the $SU(4)$ point
$(x,y)=(1/4,1/4)$,\cite{PSK,L,I,SS} which is not the focus of
this paper. In the following we apply the bond
operators in Eq. (\ref{cd2}) for both spin $S$ and pseudospin $T$ in Eq. (%
\ref{hm}).

There are four combinations of uniform FM and AFM phases: (1) $S$-FM $%
(J_{S}=y+\Theta_{F}^{2}<0)$, $T$-FM $(J_{T}=x+\Delta_{F}^{2}<0)$; (2) $S$-FM
$(J_{T}=y-\Theta_{A}^{2}<0)$, $T$-AFM $(J_{S}=x+\Delta_{F}^{2}>0)$; (3) $S$%
-AFM $(J_{T}=y+\Theta_{F}^{2}>0)$, $T$-FM$\
(J_{T}=x-\Delta_{A}^{2}<0)$; (4) $S$-AFM
$(J_{T}=y-\Theta_{A}^{2}>0)$, $T$-AFM
$(J_{S}=x-\Delta_{A}^{2}>0)$. The inequalities in the parentheses
should be fulfilled when the mean fields are solved. The four
phases correspond to four separate regions in $(x,y)$ plane. We
omit details for these uniform phases. Let us focus on more
interesting cases. It was pointed out by Shen \textit{et al.}
that OVB state should be a preferred ground state
around the point $O$: $%
(x,y)=(1,1/4)$ for $S=1$ and $T=1/2$.\cite{SXZ} This dimerization
effect is also
confirmed by the TMRG calculation of orbital-orbital correlation function,%
\cite{SK} and is also thought to be responsible for the observation on the
neutron scattering experiment of YVO$_{3}$.\cite{U} In fact, the generalized
model of Eq. (\ref{h}) provides two other valence bond solid states, the
spin valence bond solid (SVB) and spin-orbital valence bond solid (SOVB).
The three kinds of dimer phases have not been studied and distinguished
within an unified framework in the previous works. To approach the dimer
phases, we has two schemes as depicted in Fig. 1(a) and (b). It turns out
that the scheme (b) is only appropriate for exploring the SOVB phase and
does not produce the OVB and SVB phases. So we shall focus on the scheme (a).

\begin{figure}[ptb]
\begin{center}
\includegraphics[width=8.0cm]{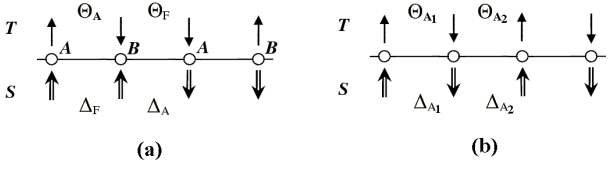}
\end{center}
\caption{Two schemes to approach the dimer phases. The single and
double arrow lines represent T and S, respectively. Two parallel
arrows represent FM channel, and two antiparallel arrows
represent AFM channel. }
\end{figure}

One can find, for the scheme (a), the spin $S$ chain and the pseudospin $T$
chain are the same object mathematically: \textit{dimerized chain with
alternating FM and AFM bonds}. For the spin $S$ part, we can divide it into
two sublattices $A$ and $B$ by assuming that the translational invariance is
broken simultaneously. In fact, the period length of the chain is doubled
due to the spin-Peierles phenomenon. We use two sets of Schwinger bosons $%
\left( a,b\right) $ and $\left( c,d\right) $ to represent spins on the
sublattices $A$ and $B$ respectively and choose the unit cell to be composed
of two lattice sites with one for sublattice $A$ and another for sublattice $%
B$. Then the Fourier transformations read

\end{subequations}
\begin{subequations}
\begin{align}
\binom{a_{j}}{b_{j}} & =\frac{1}{\sqrt{N}}\sum_{k}e^{ikR_{j}}\binom{a_{k}}{%
b_{k}}, \\
\binom{c_{j+\delta}}{d_{j+\delta}} & =\frac{1}{\sqrt{N}}\sum_{k}e^{ik\left(
R_{j}+\delta\right) }\binom{c_{k}}{d_{k}}, \\
R_{j} & =j,\ \delta=\frac{1}{2}.  \label{FT}
\end{align}
Supposed nonzero real mean fields are

\end{subequations}
\begin{subequations}
\begin{align}
\Delta_{A} & =\left\langle \frac{1}{2}\left(
a_{j}d_{j+\delta}-b_{j}c_{j+\delta}\right) \right\rangle , \\
\Delta_{F} & =\left\langle \frac{1}{2}\left(
a_{j}^{\dag}c_{j-\delta}+b_{j}^{\dag}d_{j-\delta}\right) \right\rangle .
\end{align}
The Lagrangian multipliers, $\lambda_{j,S}^{(A)}$ and $\lambda_{j,S}^{(B)}$,
are imposed to realize the constraints on the spin $S$ by adding the term in
the total Hamiltonian,

\end{subequations}
\begin{align}
& \sum_{j}\lambda_{j,S}^{(A)}\left(
a_{j}^{\dag}a_{j}+b_{j}^{\dag}b_{j}-2S\right)  \notag \\
& +\sum_{j}\lambda_{j,S}^{(B)}\left( c_{j+\delta}^{\dag}c_{j+\delta
}+d_{j+\delta}^{\dag}d_{j+\delta}-2S\right) .
\end{align}
For a mean field approach we take $\lambda_{j,S}^{(A)}=\lambda_{j,S}^{(B)}=%
\lambda_{S}.$ After the Fourier transform, we introduce the Nambu spinor in
the momentum space, $\phi_{k}^{\dagger}=\left( a_{k}^{\dagger
},b_{-k},c_{k}^{\dagger},d_{-k},a_{-k},b_{k}^{\dagger},c_{-k},d_{k}^{\dagger
}\right) .$ One can arrive at the mean-field Hamiltonian for the spin $S$
chain in\ compact form,

\begin{equation}
H^{S}=\frac{1}{2}\sum_{k}\phi_{k}^{\dagger}M^{S}\phi_{k}+\varepsilon_{0}^{S},
\end{equation}
where

\begin{equation}
\varepsilon_{0}^{S}=N\left[ \left( \Theta_{F}^{2}+y\right)
\Delta_{A}^{2}+\left( \Theta_{A}^{2}-y\right) \Delta_{F}^{2}\right]
-2N\lambda _{S}(2S+1),
\end{equation}
and the matrix $M^{S}$ is constructed by the Kronecker products of the Pauli
matrices,

\begin{align}
M^{S}& =\lambda _{S}\sigma _{0}\otimes \sigma _{0}\otimes \sigma _{0}  \notag
\\
& +j_{A,S}\cos \frac{k}{2}\ \sigma _{0}\otimes \sigma _{y}\otimes \sigma
_{y}+j_{A,S}\sin \frac{k}{2}\ \sigma _{0}\otimes \sigma _{x}\otimes \sigma
_{y}  \notag \\
& -j_{F,S}\cos \frac{k}{2}\ \sigma _{0}\otimes \sigma _{x}\otimes \sigma
_{z}+j_{F,S}\sin \frac{k}{2}\ \sigma _{0}\otimes \sigma _{y}\otimes \sigma
_{z}
\end{align}%
with $j_{A,S}=\left( \Theta _{F}^{2}+y\right) \Delta _{A}/2,\ j_{F,S}=\left(
\Theta _{A}^{2}-y\right) \Delta _{F}/2,\ j_{A,T}=\left( \Delta
_{F}^{2}+x\right) \Theta _{A}/2,\ $and $j_{F,T}=\left( \Delta
_{A}^{2}-x\right) \Theta _{F}/2$. Then we can read out the spectra from the
poles of the bosonic Matsubara Green's function ($8\times 8$ matrix),

\begin{equation}
G^{S}\left( k,i\omega_{n}\right) =\left( i\omega_{n}\sigma_{z}\otimes
\sigma_{0}\otimes\sigma_{z}-M^{S}\right) ^{-1}  \label{Green}
\end{equation}
where $\omega_{n}=n\pi/\beta$ ($n$ is an integer and $\beta$ is the inverse
temperature). Similarly we repeat the mean field calculation on the
pseudospin $T$ chain. Both the spin and the pseudospin chains have two
spectra, which read

\begin{equation}
\omega_{\alpha,\nu}\left( k\right) =\sqrt{\lambda_{\alpha}^{2}-j_{A,\alpha
}^{2}+j_{F,\alpha}^{2}+2\nu j_{F,\alpha}\Omega_{\alpha}\left( k\right) },
\label{spectra}
\end{equation}
where $\alpha=(S,T)$, $\nu=\mp$, and $\Omega_{\alpha}\left( k\right) =\sqrt{%
\lambda_{\alpha}^{2}-j_{A,\alpha}^{2}\cos^{2}k}$. The lower quasiparticle
spectrum $\omega_{\alpha,-}\left( k\right) $ is found to be gapped at $%
k=\pm\pi/2$, which is related to the physical gap by the peak of the
imaginary part of the dynamic spin susceptibility.\cite{A} And the gap is
also true for half integer spin (or pseudospin).\cite{SXZ,MYK} By optimizing
the total free energy

\begin{equation}
F=\varepsilon_{0}-\frac{1}{\beta}\sum\limits_{k,\alpha,\nu}\ln\left[
n_{B}\left( \omega_{\alpha,\nu}\left( k\right) \right) \left\{ n_{B}\left(
\omega_{\alpha,\nu}\left( k\right) \right) +1\right\} \right] ,
\end{equation}
where $n_{B}(\omega)$ is the Bose-Eienstein distribution and

\begin{align}
\varepsilon_{0} & =N[3\left(
\Theta_{A}^{2}\Delta_{F}^{2}+\Theta_{F}^{2}\Delta_{A}^{2}\right) +2xy+\left(
\Delta_{A}^{2}-\Delta_{F}^{2}\right) y  \notag \\
& +\left( \Theta_{A}^{2}-\Theta_{F}^{2}\right)
x-2\lambda_{S}(2S+1)-2\lambda_{T}(2T+1)],
\end{align}
a set of the mean field equations are established,

\begin{subequations}
\begin{align}
W\left( \widetilde{j}_{A,S},\widetilde{j}_{F,S}\right) & =2S+1, \\
W\left( \widetilde{j}_{A,T},\widetilde{j}_{F,T}\right) & =2T+1, \\
X\left( \widetilde{j}_{A,S},\widetilde{j}_{F,S}\right) & =\Delta_{A}, \\
X\left( \widetilde{j}_{A,T},\widetilde{j}_{F,T}\right) & =\Theta_{A}, \\
Y\left( \widetilde{j}_{A,S},\widetilde{j}_{F,S}\right) & =\Delta_{F}, \\
Y\left( \widetilde{j}_{A,T},\widetilde{j}_{F,T}\right) & =\Theta_{F}, \\
\frac{\left( \Theta_{F}^{2}+y\right) \Delta_{A}}{\widetilde{j}_{A,S}} & =%
\frac{\left( \Theta_{A}^{2}-y\right) \Delta_{F}}{\widetilde{j}_{F,S}}%
=2\lambda_{S}, \\
\frac{\left( \Delta_{F}^{2}+x\right) \Theta_{A}}{\widetilde{j}_{A,T}} & =%
\frac{\left( \Delta_{A}^{2}-x\right) \Theta_{F}}{\widetilde{j}_{F,T}}%
=2\lambda_{T},  \label{mfe}
\end{align}
where we have defined integrals (the symbols with wave decoration means
dimensionless quantities, $\widetilde{j}_{A,\alpha}=j_{A,\alpha}/\lambda_{%
\alpha},\widetilde{j}_{F,\alpha}=j_{F,\alpha}/\lambda_{\alpha },\widetilde{%
\Omega}_{\alpha}=\Omega_{\alpha}/\lambda_{\alpha}$, and we have substituted
the sum with integral: $\frac{1}{N}$ $\sum_{k}\rightarrow\int \frac{dk}{2\pi}
$),

\end{subequations}
\begin{subequations}
\begin{align}
W(\widetilde{j}_{A,\alpha },\widetilde{j}_{F,\alpha })& =\int \frac{dk}{2\pi
}\sum_{\nu =\mp }\left( 1+\frac{\nu \widetilde{j}_{F,\alpha }}{\widetilde{%
\Omega }_{\alpha }}\right) \frac{\coth \frac{\beta \omega _{\alpha ,\nu }}{2}%
}{2\widetilde{\omega }_{\alpha ,\nu }}, \\
X(\widetilde{j}_{A,\alpha },\widetilde{j}_{F,\alpha })& =\int \frac{dk}{2\pi
}\sum_{\nu =\mp }\left( 1+\frac{\nu \widetilde{j}_{F,\alpha }\cos ^{2}k}{%
\widetilde{\Omega }_{\alpha }}\right) \frac{\coth \frac{\beta \omega
_{\alpha ,\nu }}{2}}{4\widetilde{\omega }_{\alpha ,\nu }}, \\
Y(\widetilde{j}_{A,\alpha },\widetilde{j}_{F,\alpha })& =-\int \frac{dk}{%
2\pi }\sum_{\nu =\mp }\left( \widetilde{j}_{F,\alpha }+\nu \widetilde{\Omega
}_{\alpha }\right) \frac{\coth \frac{\beta \omega _{\alpha ,\nu }}{2}}{4%
\widetilde{\omega }_{\alpha ,\nu }}.
\end{align}

\section{Phase diagram}

\begin{figure}[ptb]
\begin{center}
\includegraphics[width=8.0cm]{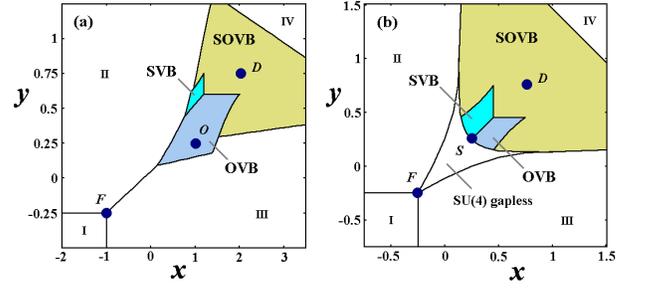}
\end{center}
\caption{(Color online) Phase diagrams for: (a) $S=1,T=1/2$; (b) $%
S=1/2,T=1/2 $. The uniform phases are:\ I for S-FM and T-FM; II for S-AFM
and T-FM; III for S-FM and T-AFM; IV for S-AFM and T-AFM. The SU(4) gapless
region is quoted from Ref. \protect\cite{I} The special points are: $F$ -
the FM point; $D$ - the dimer point, $S$ - the SU(4) symmetric point, and $O$
- the representative point of OVB phase. Please see details in the text.}
\end{figure}

For the transition-metal oxides, the pseudospin is usually one half, i. e. $%
T=1/2$, reflecting two choices of unfrozen $e_{g}$ or $t_{2g}$ orbitals.\cite%
{TN,KHO} Spin-gap materials, Na$_{2}$Ti$_{2}$Sb$_{2}$O and NaV$_{2}$O$_{5}$,
are related to the $S=1/2$ case in the model of Eq. (\ref{h}),\cite%
{PSK,Fujii} while cubic vanadates, LaVO$_{3}$ and YVO$_{3}$, are related to $%
S=1$ case due to large Hund's coupling.\cite{KHO,MFM}

The phase diagrams are obtained by solving the mean field equations
numerically at zero temperature. Fig. 2(a) shows the case for $S=1$ and $%
T=1/2$. We found three dimer phases: (1) the OVB phase with $\Theta_{A}=%
\sqrt{3}/2,\Theta_{F}=0,\Delta_{A}\neq0$ and $\Delta_{F}\neq0$; (2) the SVB
phase with $\Theta_{A}\neq0,\Theta_{F}\neq0,\Delta_{A}=\sqrt{2}$ and $%
\Delta_{F}=0$; (3) the SOVB phase with $\Theta_{A}=\sqrt{3}/2,\Theta
_{F}=0,\Delta_{A}=\sqrt{2}$ and $\Delta_{F}=0$. As expected, the point $O$: $%
(x,y)=(1,1/4)$ lies in the OVB phase region. At the mean field level, the
dimer phases are captured by the staggered non-zero AFM mean fields $%
\Theta_{A}$ and/or $\Delta_{A}$, which provide a perfect dimer picture of
spin and/or pseudospin in the whole phase regions. The reason for this maybe
lies in two: the mean field treatment omits part of quantum fluctuations;
the dimerization effect is very strong in such a system at least at zero
temperature. In fact, TMRG has confirmed the orbital dimerization at the
point $O$: $\left( x,y\right) =\left( 1,1/4\right) $ since the correlation
function $\left\langle \mathbf{T}_{m}\cdot\mathbf{T}_{m+1}\right\rangle $
extrapolates to $-3/8$ per bond at zero temperature, although this point is
not exactly soluble.\cite{SK} And the OVB phase is quite robust even when
the anisotropy, Hund's coupling, and atomic spin-orbit interaction are taken
into account for cubic vanadates and relevant systems.\cite{SXZ,HKO} We also
guess these dimer phases may survive in two dimensions, although no rigorous
soluble point can be referred to. A dimerized OVB configuration in two
dimensions has been proposed and checked by the spin-wave theory in Ref.
\cite{SXZ}, which indeed exhibits lower energy than the uniform phases.
However it is an interesting problem whether SOVB can survive in two
dimensions.

Fig. 2(b) shows the case for $S=1/2$ and $T=1/2$. In Fig. 2(b), the $SU(4)$
gapless region is quoted from Ref. \cite{I}. We found the SVB and OVB phases
still occupy two areas that are sandwiched between the SOVB and $SU(4)$
gapless region. The three dimer phases with corresponding mean fields are:
(1) the OVB phase with $\Theta_{A}=\sqrt{3}/2,\Theta_{F}=0,\Delta_{A}\neq0$
and $\Delta_{F}\neq0$; (2) the SVB phase with $\Theta_{A}\neq0,\Theta_{F}%
\neq0,\Delta_{A}=\sqrt{3}/2$ and $\Delta_{F}=0$; (3) the SOVB phase with $%
\Theta_{A}=\sqrt{3}/2,\Theta_{F}=0,\Delta_{A}=\sqrt{3}/2$ and $\Delta_{F}=0$%
. Previous works ignored the SVB and OVB phases.\cite{PSK,I} And there is a $%
S$-AFM and $T$-AFM phase region from the point of view of SBMFT.

The result here in fact suggests that the spin-orbit coupling of
$SU(2)\times SU(2)$ type plays the roles of the intrinsic
spin-Peierls and orbital-Peierls mechanisms. The SVB formation in
this $1D$ model is an example that expresses the concept of
\textit{orbitally driven Peierls state}.\cite{Khomskii} But
notice we did not consider the lattice distortion here. For spin
$S=1$, the SVB phase is harder to realize since the region
shrinks.

\begin{figure}[ptb]
\begin{center}
\includegraphics[width=7.50cm]{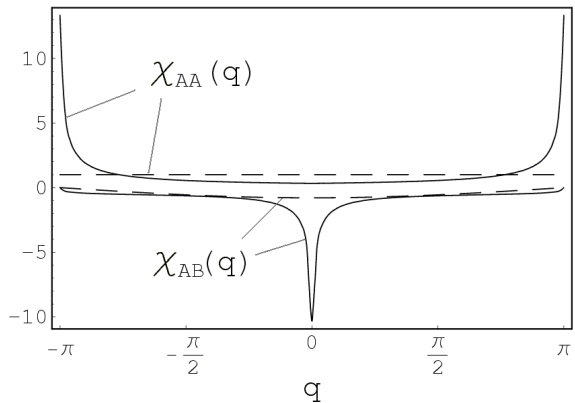}
\end{center}
\caption{Static spin susceptibilities for $S=1$ and $T=1/2$ at the OVB point
$O$: $(x,y)=(1,1/4)$ (solid line) and the dimer point $D$: $(2,3/4)$ (dashed
line).}
\end{figure}

\section{Susceptibility}

The spin structures of the dimer phases are detectable by spin
susceptibility. Let us consider the spin susceptibility (the psedospin case
is handled in the same way). We can define the spin density waves on
sublattices $A$ and $B$ (due to the rotational symmetry, we only consider $z$
component of the spin),

\end{subequations}
\begin{subequations}
\begin{align}
S_{A}^{z}\left( q\right) & =\frac{1}{\sqrt{N}}\sum e^{-iqR_{j}}S_{A,j}^{z},
\\
S_{B}^{z}\left( q\right) & =\frac{1}{\sqrt{N}}\sum e^{-iq\left(
R_{j}+\delta\right) }S_{B,j+\delta}^{z}.
\end{align}
The spin susceptibility contains contributions from both the
intra-sublattice and inter-sublattice fluctuations,

\end{subequations}
\begin{align}
\chi\left( q\right) & =\frac{1}{N}\sum_{m,n\in A,B}\left\langle
S_{m}^{z}S_{n}^{z}\right\rangle e^{iq(R_{m}-R_{n})}  \notag \\
& =\chi_{AA}\left( q\right) +\chi_{AB}\left( q\right) +\chi_{BA}\left(
q\right) +\chi_{BB}\left( q\right) .
\end{align}
In Matsubara formalism, the intra-sublattice and inter-sublattice
contributions read

\begin{subequations}
\begin{align}
\chi_{AA}\left( q,\tau\right) & =\left\langle S_{A}^{z}\left( q,\tau\right)
S_{A}^{z}(-q,0)\right\rangle  \notag \\
& =\frac{1}{4}\int_{-\pi}^{\pi}\frac{dk}{2\pi}[2G_{55}^{S}\left(
-k,\tau\right) G_{11}^{S}\left( k+q,\tau\right)  \notag \\
& \qquad\qquad-G_{56}^{S}\left( -k,\tau\right) G_{12}^{S}\left(
k+q,\tau\right)  \notag \\
& \qquad\qquad-G_{21}^{S}\left( -k,\tau\right) G_{65}^{S}\left(
k+q,\tau\right) ], \\
\chi_{AB}\left( q,\tau\right) & =\left\langle S_{A}^{z}\left( q,\tau\right)
S_{B}^{z}(-q,0)\right\rangle  \notag \\
& =\frac{1}{4}\int_{-\pi}^{\pi}\frac{dk}{2\pi}[2G_{57}^{S}\left(
-k,\tau\right) G_{13}^{S}\left( k+q,\tau\right)  \notag \\
& \qquad\qquad-G_{58}^{S}\left( -k,\tau\right) G_{14}^{S}\left(
k+q,\tau\right)  \notag \\
& \qquad\qquad-G_{23}^{S}\left( -k,\tau\right) G_{67}^{S}\left(
k+q,\tau\right) ],
\end{align}
respectively, where the Green's function in imaginary time $\tau$ is related
to Eq. (\ref{Green}) by

\end{subequations}
\begin{equation}
G^{S}\left( k,\tau\right) =\frac{1}{\beta}\sum_{n}e^{-i\omega_{n}\tau}G^{S}%
\left( k,i\omega_{n}\right) .
\end{equation}

The numerical solutions of the static spin susceptibilities, $\chi
_{AA}\left( q,\tau =0^{+}\right) $ and $\chi _{AB}\left( q,\tau
=0^{+}\right) $, for $S=1$ and $T=1/2$ at the OVB point $O$: $(x,y)=(1,1/4)$
and the dimer point $D$: $(2,3/4)$ are shown in Fig. 3. At the OVB point $O$%
,\ the static spin susceptibilities exhibit finite sharp peaks at $q=\pi $
for $\chi _{AA}\left( q,\tau \right) $ and $q=0$ for $\chi _{AB}\left(
q,\tau \right) $ reflecting the strong AFM fluctuation among sites in
sublattice $A$ (or $B$) and FM fluctuation between sites in sublattice $A$
and sites in sublattice $B$. The finite peaks also indicate no long-range
order exists at zero temperature. While there is no sharp peak at the dimer
point $D$, since the formation of spin singlets leads to dispersionless spin
spectra, $\omega _{S,-}\left( k\right) =\omega _{S,+}\left( k\right) =\sqrt{%
\lambda _{S}^{2}-j_{A,S}^{2}}$. The spin susceptibilities at the OVB point $O
$ and the dimer point $D$ characterize the whole OVB region and SOVB region
respectively. As we go from OVB phase to SVB phase, the spin and pseudospin
change their roles. In SOVB phase, the sharp peaks will disappear for both
spin and pseudospin. Thus the three dimer phases are well characterized by
the spin and pseudospin susceptibilities and distinguishable experimentally.
A recent neutron scattering experiment of YVO$_{3}$ has provided a tentative
proof of the OVB.\cite{U}

\section{Conclusion}

In conclusion, we have employed SBMFT to study the valence bond states as
well as uniform FM and AFM states in the $1D$ $SU(2)\times SU(2)$
spin-orbital model. The phase diagrams are constructed by comparing the
ground energies of the proposed possible states. In the valence bond states
the translational invariance is broken in space, FM and AFM parameters
compete with each other and are determined self-consistently. Specifically,
two sublattices are introduced to approach the three dimerized SVB, OVB and
SOVB phases by SBMFT. We clarified how to capture properly the low
temperature properties of such a system by SBMFT. The main result of the
paper is that the SVB and OVB have lower energy than the SOVB\ in some
regions. These consequences reveal that the $SU(2)\times SU(2)$ type
spin-orbit coupling can provide both spin-Peierls and orbital-Peierls
mechanisms. The results still should be compared with other phases including
those with translational invariance. Static spin and pseudospin
susceptibilities had been calculated in the present theory and are available
to distinguish spin-singlet and orbital-singlet formations in real materials.

This work was supported by the Research Grant Council of Hong Kong (No.:
HKU/7109/02P and 7038/04P)


\begin{thebibliography}{99}
\bibitem{TN} Y. Tokura and N. Nagaosa, Science \textbf{288}, 462 (2000).

\bibitem{KK} I. Kugel and D. I. Khomskii, Sov. Phys. JETP lett. \textbf{37},
725 (1973); Sov. Phys. Usp. \textbf{25}, 231 (1982).

\bibitem{R} T. M. Rice, \textit{Spectroscopy of Mott Insulators and
Correlated Metals}, edited by A. Fujimori and Y. Tokura (Springer, Berlin,
1995).

\bibitem{AA} D. P. Arovas and A. Auerbach, Phys. Rev. B \textbf{52}, 10114
(1995).

\bibitem{PSK} S. K. Pati, R. R. P. Singh and D. I. Khomskii, Phys. Rev.
Lett. \textbf{81}, 5406 (1998).

\bibitem{Fujii} Y. Fujii, H. Nakao, T. Yosihama, M. Nishi, K. Nakajima, K.
Kakurai, M. Isobe, Y. Ueda and H. Sawa, J. Phys. Soc. Jpn. \textbf{66}, 326
(1997).

\bibitem{KHO} G. Khaliullin, P. Horsch and A. M. Oles, Phys. Rev. Lett.
\textbf{86}, 3879 (2001).

\bibitem{I} C. Itoi, S. Qin and I. Affleck, Phys. Rev. B \textbf{61}, 6747
(2000).

\bibitem{ZO} W. Zheng and J. Oitmaa, Phys. Rev. B \textbf{64}, 014410 (2001).

\bibitem{K} A. K. Kolezhuk and H. -J. Mikeska, Phys. Rev. Lett. \textbf{80},
2709 (1998); K. Itoh, J. Phys. Soc. Jpn. \textbf{68}, 322 (1999).

\bibitem{L} Y. Q. Li, M. Ma, D. N. Shi and F. C. Zhang, Phys. Rev. Lett.
\textbf{81}, 3527 (1998).

\bibitem{SXZ} S. Q. Shen, X. C. Xie and F. C. Zhang, Phys. Rev. Lett.
\textbf{88}, 27201 (2002).

\bibitem{SK} J. Sirker and G. Khaliullin, Phys. Rev. B \textbf{67}, 100408
(2003).

\bibitem{U} C. Ulrich, G. Khaliullin, J. Sirker, M. Reehuis, M. Ohl, S.
Miyaska, Y. Tokura and B. Keimer, Phys. Rev. Lett. \textbf{91}, 257202
(2003).\newline

\bibitem{A} A. Auerbach, \textit{Interacting Electrons and Quantum Magnetism}
(Springer-Verlag, New York, 1994).

\bibitem{AA1} A. Auerbach and D. Arovas, Phys. Rev. Lett. \textbf{61}, 617
(1988); Phys. Rev. B. \textbf{38}, 316 (1988).

\bibitem{S} S. Sarker, C. Jayaprakash, H. R. Krishnamurthy and M. Ma, Phys.
Rev. B. \textbf{40}, 5028 (1989).

\bibitem{Zhang01prl} G. M. Zhang and S. Q. Shen, Phys. Rev. Lett. 87,
157201(2001); S. Q. Shen and G. M. Zhang, Europhys. Lett. 57, 274 (2002).

\bibitem{C} H. A. Ceccatto, C. J. Gazza and A. E. Trumper, Phys. Rev. B
\textbf{47}, 12329 (1993).

\bibitem{T} A. E. Trumper, L. O. Manuel, C. J. Gazza and H. A. Ceccatto,
Phys. Rev. Lett. \textbf{78}, 2216 (1997).

\bibitem{SS} S. Q. Shen, Phys. Rev. B \textbf{66}, 214516 (2002); P. Li and
S. Q. Shen, N. J. Phys. \textbf{6}, 160 (2004).

\bibitem{MYK} H. Manaka, I. Yamada, T. Kikuchi, K. Morishita and K. Iio, J.
Phys. Soc. Jpn. \textbf{70}, 2509 (2001).\newline

\bibitem{MFM} T. Mizokawa and A. Fujimori, Phys. Rev. B \textbf{54}, 5368
(1996); F. Mila, R. Shiina, F. C. Zhang, A. Joshi, M. Ma, V. Anisimov and T.
M. Rice, Phys. Rev. Lett. \textbf{85}, 1714 (2000).\newline

\bibitem{HKO} P. Horsch, G. Khaliullin and A. M. Oles, Phys. Rev. Lett.
\textbf{91}, 257203 (2003).\newline

\bibitem{Khomskii} D. I. Khomskii and T. Mizokawa, Phys. Rev. Lett. \textbf{%
94}, 156402 (2005).\newline
\end{thebibliography}
\end{document}